\documentclass[%
 aps,
 superscriptaddress,
 amsmath,amssymb,
reprint,%
 numerical
]{revtex4-1}

\usepackage[utf8]{inputenc}
\usepackage[T1]{fontenc}

\usepackage{graphicx}
\usepackage{hyperref}
\usepackage{float}
\usepackage[dvipsnames]{xcolor}
\colorlet{green1}{green!40!black!100!}
\usepackage{dcolumn}
\usepackage{bm}
\usepackage{mathptmx}
\usepackage{etoolbox}
\usepackage{relsize}

\makeatother

\begin{document}

\preprint{APS/123-QED}

\title{Modified Poisson-Boltzmann theory for polyelectrolytes in monovalent salt solutions with finite-size ions}

\author{Hossein Vahid}
\affiliation{Department of Chemistry and Materials Science, Aalto University, P.O. Box 16100, FI-00076 Aalto, Finland}
\affiliation{Department of Applied Physics, Aalto University, P.O. Box 11000, FI-00076 Aalto, Finland}

\author{Alberto Scacchi}
\affiliation{Department of Chemistry and Materials Science, Aalto University, P.O. Box 16100, FI-00076 Aalto, Finland}
\affiliation{Department of Applied Physics, Aalto University, P.O. Box 11000, FI-00076 Aalto, Finland}

\author{Xiang Yang}
\affiliation{Department of Applied Physics, Aalto University, P.O. Box 11000, FI-00076 Aalto, Finland}
\affiliation{Quantum Technology Finland Center of Excellence, Department of Applied Physics, Aalto University, P.O. Box 11000, FI-00076 Aalto, Finland}

\author{Tapio Ala-Nissila}
\affiliation{Quantum Technology Finland Center of Excellence, Department of Applied Physics, Aalto University, P.O. Box 11000, FI-00076 Aalto, Finland}
\affiliation{Interdisciplinary Centre for Mathematical Modelling and Department of Mathematical Sciences, Loughborough University, Loughborough, Leicestershire LE11 3TU, UK}

\author{Maria Sammalkorpi}\email{maria.sammalkorpi@aalto.fi}
\affiliation{Department of Chemistry and Materials Science, Aalto University, P.O. Box 16100, FI-00076 Aalto, Finland}
\affiliation{Department of Bioproducts and Biosystems, Aalto University, P.O. Box 16100, FI-00076 Aalto, Finland}
\affiliation{Academy of Finland Center of Excellence in Life-Inspired Hybrid Materials (LIBER), Aalto University, P.O. Box 16100, FI-00076 Aalto, Finland}

\date{\today}

\begin{abstract}

We present a soft-potential-enhanced Poisson-Boltzmann (SPB) theory to efficiently capture
ion distributions and electrostatic potential around rodlike charged macromolecules. The SPB model is calibrated with a coarse-grained particle-based model for polyelectrolytes (PEs) in monovalent salt solutions as well as compared to a full atomistic molecular dynamics simulations with explicit solvent. We demonstrate that our modification enables the SPB theory to accurately predict monovalent ion distributions around a rodlike PE in a wide range of ion and charge distribution conditions in the weak-coupling regime. These include excess salt concentrations up to $1$~M, and ion sizes ranging from small ions, such as Na$^+$ or Cl$^-$, to softer and larger ions with size comparable to the PE diameter. The work provides a simple way to implement an enhancement that effectively captures the influence of ion size and species into the PB theory in the context of PEs in aqueous salt solutions.
\end{abstract}

\maketitle

\section{Introduction}

Solutions and assemblies of  polyelectrolytes (PEs), polymers with electrolyte groups that dissociate in aqueous environment, are ubiquitous in biology, but also in industrial processes and advanced materials. The applications of PEs range from hydrogels and water purification, synthetic biology and drug
transport to flocculation, bio-contact materials, and
chemical sensors in aqueous environments~\cite{boudou2010, delcea2011, stuart2010, rivas2011, zhang2018}. In the applications, PE interactions and assembly are readily controlled by, e.g., PE chemistry, pH, temperature, and solution additives. A particularly important and simple way to control the interactions is the addition of salt, see e.g. Refs. \onlinecite{szilagyi2014, becker2012}. 

To a first approximation, the PEs, such as many synthetic and biopolymers, including e.g. DNA and many proteins, can be approximated as charged rods or cylinders at scales well below the persistence length.
An accessible and relatively simple means to assess interactions of rod-like charged macromolecules in solution is provided by the mean-field Poisson–Boltzmann (PB) description, in which the charged macromolecule is described as an idealized, infinitely long line charge
in the weak-coupling regime (for reviews see e.g. Refs.~\onlinecite{Ren2012,grochowski2008}). Based on the PB model, the most well-known approach to consider ion condensation in PE systems is provided by the Manning-Oosawa theory that describes the counterions as pointlike and considers them within the Debye–H\"uckel approximation \cite{manning1969,oosawa1968}. Manning description has also been augmented to cover excess salt via the nonlinear Poisson–Boltzmann model \cite{sharp1995}. 

Characteristic to the mean-field approaches, such as the PB theory, is that it captures the PE-system response at low charge densities and for monovalent salt solutions~\cite{deserno2000paper,naji2006s,buyukdagli2011}. However excess multivalent salt or strongly interacting systems break down PB based approaches~\cite{deserno2000paper,trizac2006,deserno2001,buyukdagli2016,perel1999,buyukdagli2017}. To describe these, approaches going beyond the weak-coupling PB equation, strong-coupling theory approaches~\cite{gronbech1997,rouzina1996,grosberg2002} or particle-based simulations~\cite{deserno2000paper,deserno2001} need to be used.
Specifically, mean-field models fail to account for molecular level structure and  PE charge distribution deviations from the assumed cylindrical symmetry \cite{Ren2012,cisneros2014}. Ion size can be effectively accounted for in the models~\cite{borukhov1997,lopez2011,zhou2011,colla2017} and also chemical specificity of the ions can be partially considered for by modifying the interaction potentials, see e.g. Refs.~\onlinecite{heyda,bostrom2002,batys2017,kirmizialtin2012}. These type of enhancements significantly increase the applicability range of the PB approaches.

Particle-based Monte Carlo and molecular dynamics simulations can efficiently capture the structural and charge-correlations-related features that mean-field approaches fail with. Previously, particle-based simulations~\cite{deserno2000paper,nguyen2000,Antila2016,antila2017} 
of PE systems have outlined the significance of charge correlations in PE interactions. Furthermore, it is obvious that at microscopic lengths scales the molecular structure is important, see e.g. Ref. \onlinecite{batys2017}.

Nevertheless, due to their conceptual simplicity and numerical efficiency, 
enhancement schemes to PB models remain very useful for describing PE interactions. Notably, the modified mean-field approaches are parameter-dependent in accuracy and may involve several input parameters with no standardized way of determining them, see e.g. Refs. \onlinecite{dong2003, pang2013, harris2013}.  Here, we target an easy-to-approach effective PB theory to enhanced modelling of PEs in aqueous salt solutions and a systematic modelling accuracy characterization. To this purpose, we consider a PE in aqueous salt solution as a charged rod, its counterions, and the monovalent salt as added ions. We construct a soft-potential-enhanced Poisson-Boltzmann (SPB) model, parametrize it against particle-based simulations and systematically map the accuracy of the description against the particle based predictions in coarse-grained and full atomistic detail levels. The model is thus based on a rigorous description of the system and the implemented modification has a physical basis. The work demonstrates an easily parametrizable, single-parameter SPB model that captures to a very good degree of accuracy ion distributions and the electrostatic potential around the examined model PE in monovalent salt. This includes concentrations ranging from no added salt to $1$ M, and ion sizes corresponding to small, hard monovalent ions such as Na$^+$ to significantly larger, spherical ions with hydrated effective diameter exceeding $1$ nm. 
\section{Theory and Simulation Methods}
\subsection{All-atom-detail molecular dynamics  simulations}
For the atomistic-detail reference system, we choose the anionic poly(styrene sulfonate) (PSS) with sodium counterions and NaCl as the added monovalent salt. This is because of its relatively high line charge density ($-3.6$ e/nm) and bottle-brush-like structure in which the charged sulfonate groups extend from the backbone asymmetrically. The localized charges away from the center axis of the polymer and their density cause deviations from the mean-field approach. 

The GROMACS package~\cite{van2005gromacs,abraham2015gromacs} is used for the atomistic-detail molecular dynamics simulations. We use the OPLS-AA~\cite{jorgensen1988} force field to model the PE and the TIP4P~\cite{jorgensen1985} model for explicit water molecules. Na$^+$ and Cl$^-$ ion models are taken from Refs.~\onlinecite{aqvist1990} and~\onlinecite{chandrasekhar1984}, respectively. 
  
A linear chain of $20$ monomers is set parallel to the $z$ axis and spanning the simulation box as an infinite chain (see Fig.~\ref{model_2}(a)). Preparation of the periodic, infinite chain follows Ref.~\onlinecite{batys2017}. Solvation is performed by the GROMACS solvate tool, and the final water-density-equilibrated simulation box is $7.9 \times 7.9 \times 5.6$~nm$^3$. All atomistic-detail simulation systems contain $20$ Na$^+$ ions as PSS counterions. We add excess salt in concentrations of $0.125$~M, $0.25$~M, $0.5$~M, and $1$~M to the solvated system by random replacement of water molecules.

A cutoff of $1$ nm is used for non-bonded interactions and electrostatic contributions in real space, whereas a direct cutoff (no shift) is used for the Lennard-Jones potential. The long-range electrostatic contributions are calculated using the particle-mesh Ewald (PME) method~\cite{essmann1995} with $0.16$~nm grid spacing and fourth-order splines. Temperature is controlled by the stochastic $V$-rescale thermostat~\cite{bussi2007} with a coupling constant of $0.1$~ps and reference temperature $T = 300$~K. Pressure control is semi-isotropic with Parrinello–Rahman barostat~\cite{parrinello,nose1983} with a coupling constant of $1$~ps and reference pressure $1$~bar. Following Ref.~\onlinecite{batys2017}, the system is set to be incompressible along the $z$ axis. A $2$ fs time step within the leap-frog integration scheme is applied in the $NpT$ simulations. Additionally, all bonds in PEs and in water molecules are constrained using the LINCS~\cite{lincs} and SETTLE~\cite{miyamoto1992} algorithms, respectively.
 
After initialization, steepest-descent energy minimization is performed. This is followed by a $2$ ns semi-isotropic $NpT$ ensemble simulation where the PSS is held fixed to adjust the $xy$ dimensions and water and ion distributions around the PE. 
For the production runs, the PSS is released such that it can freely translate in the $xy$ plane and rotate around its axis. A $100$~ns $NpT$ simulation is carried out, and the first $2$ ns are disregarded from the analysis. Data sampling is performed every $5$~ps.

\subsection{Coarse-grained detail molecular dynamics simulations}\label{sec:coarse-grained}

In coarse-grained molecular dynamics modelling, we construct a charged rod by linearly adding up spherical beads (force centers) with charge $\zeta e$ each, where $e$ is the elementary charge and $\zeta$ the charge valency. The rod beads are at equal separation $b$, which results in a line charge $\lambda = \zeta e/b$. 
The ions are spherical beads of monovalent charge $\pm e$. 

The interactions between particles are modelled via a Weeks-Chandlers-Andersen (WCA)~\cite{anderson} potential, which generally reads as
\begin{equation}
       V^{ ij}(r)= 4 \epsilon^{ ij} \left[\left(\frac{\sigma^{ ij}_{\rm }}{r^{ ij}}\right)^{12} -\left(\frac{\sigma^{ ij}_{\rm }}{r^{ ij}}\right)^6 \right]+\epsilon^{ ij}; \label{lj1} \quad r<r^{ ij}_{\rm c}.
  \end{equation}
The labels $i,j$ denote either the ionic species or the polymer beads, and $V^{ij}(r)$ the pair interaction between all the particles in the system. In the present case of monovalent salt such as NaCl, there are five distinct interaction pairs corresponding to anion-anion, cation-cation, anion-cation, anion-bead, and cation-bead cases. The bead-bead interaction is not relevant here as the relative bead positions in the rod are strictly constrained to be at $b$. $r^{ij}$ is the distance between the pair $ij$, and $\sigma^{i}$ and $\epsilon^{i}$ denote the diameter and the depth of the potential well for species $i$ (excluding the bead-bead case). For the mixed interactions $i \ne j$ we use $\sigma^{ij} =(\sigma^{i}+\sigma^{j})/2$, and the Lorentz-Berthelot (LB) mixing rule $\epsilon^{ ij}=\sqrt{\epsilon^{i} \epsilon^{j}}$.
The cutoff radii for the pair interactions are set to $r^{ij}_{\rm c}=2^{1/6}\sigma^{ij}$. 

Contributions due to electrostatics are modelled via Coulombic potentials. The pairwise interaction between two ionic species I and J with charges $\zeta^{i}e$ and $\zeta^{j}e$ is given by
\begin{equation}
      \beta e V_{\rm C}(r)= \zeta^{i} \zeta^{ j}\frac{\ell_{\rm B}}{r},
  \end{equation}
where $\beta=1/k_{\rm B}T$ for $k_{\rm B}$ the Boltzmann constant and $T$ the temperature of the system. The Bjerrum length $\ell_{\rm B}=\beta e^2/(4\pi \varepsilon)$ measures the coupling strength by specifying the distance at which two unit charges have interaction energy of $k_{\rm B}T$. In this, $\varepsilon=\varepsilon_{\rm r}\varepsilon_0$ is the effective dielectric constant, $\varepsilon_{\rm r}$ being the dielectric constant of the solvent (for water, $\varepsilon_{\rm r}=77.75$ at 300 K and 1 atm~\cite{lide2004}) and $\varepsilon_0$ the vacuum dielectric constant. 

\begin{figure}[htbp!]
 \centering
 \includegraphics[width=1.0\linewidth]{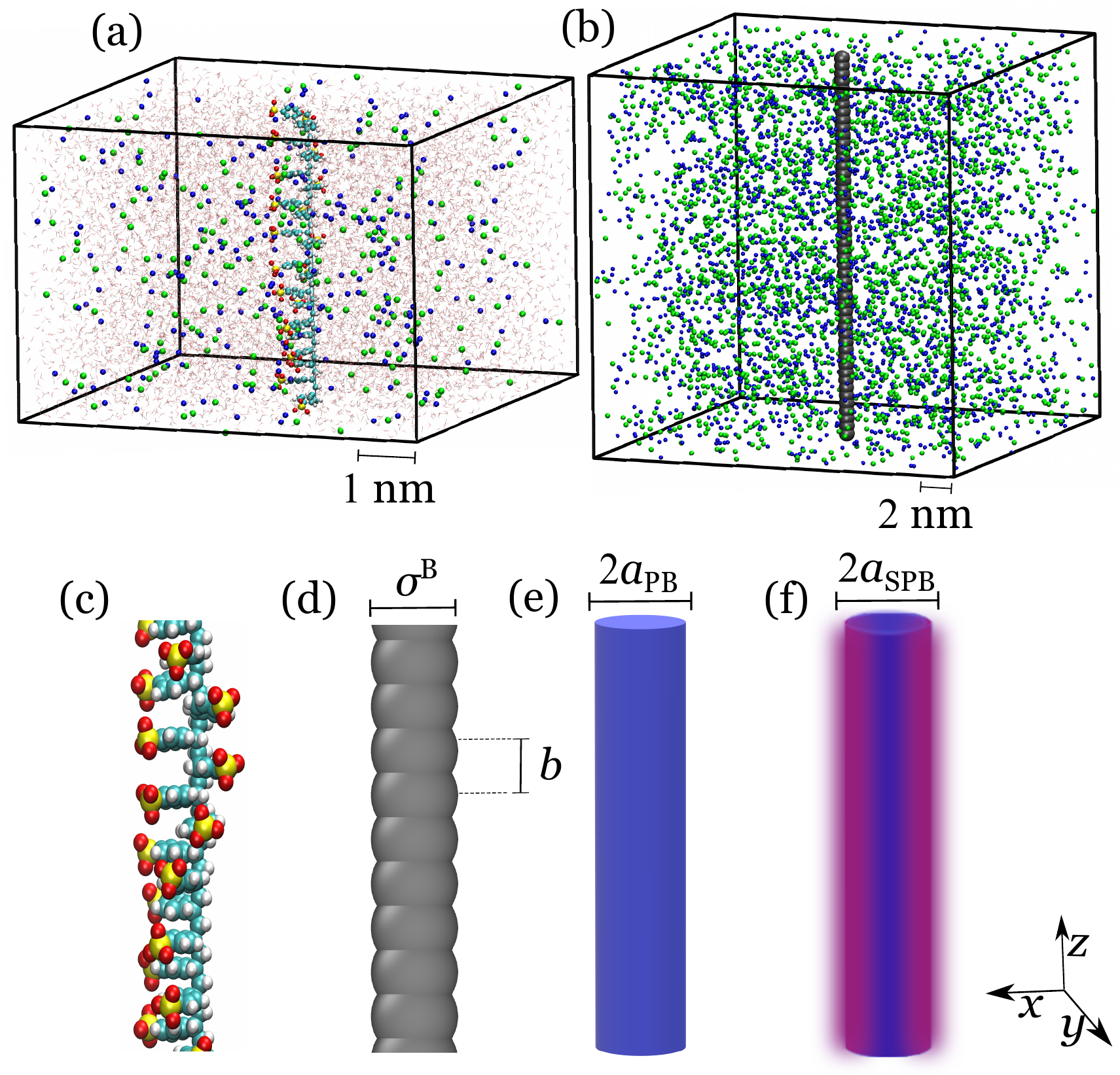}
  \caption{(a) The atomistic-detail model of PSS and (b) the coarse-grained (CG)  in periodic simulation boxes of size $7.9 \times 7.9 \times 5.6$~nm$^3$ or $20 \times 20 \times 20 $~nm$^3$ with 0.5 M salt, respectively. The explicit water molecules are present in the atomistic model while the solvent is implicit in the CG model. The Na$^{+}$ and Cl$^{-}$ ions are in blue and green, respectively, in both systems. In (c) and (d), close-up views of the atomistic-detail and CG polymer structure are presented. Panel (e) shows a schematic representation of the hard-cylinder model used in PB theories and (f) visualizes the soft-cylinder used in Eq. (\ref{density_theory3}). In (d) bead spacing $b$ and effective diameter $\sigma^{\rm B}$, in (e) the PB hard-cylinder diameter $2a_{\rm PB}$, and in (f) the PB soft-cylinder diameter $2a_{\rm SPB}$ are shown. }
 \label{model_2}
\end{figure}

The rod of beads is set along the $z$ axis into the center of a cubic simulation box of size $20\times 20\times 20$ nm$^3$ with implicit solvent and periodic boundary conditions in all directions (see Fig.~\ref{model_2}). In total, $74$ consecutive beads each with charge $\zeta e = -1 e$  are set at a distance ${b}=0.27$ nm from one another. This leads to a line charge density $\lambda = \zeta e/b = -3.6 $ e/nm, matching with the atomistic-detail PSS PE. Production simulation bead spacing ${b}=0.27$ nm was chosen such that the linear density of beads along the rod axis is increased until an essentially cylindrical object whose potential energy surfaces are uniform along the chain is obtained; see Fig.~S1 of the Supplementary Material (SM) for potential energy surfaces. The approach is consistent with our earlier work that addressed the convergence of electrostatics calculations accuracy vs. line charge discretization density~\cite{antila2015}. In checking the finite-size effects, counterion condensation and density profiles converge to system size independent values when a cubic simulation box is ($20$ nm)$^3$ or larger. To this end, cubic simulation boxes of sides $4$, $6$, $10$, $20$, $40$, and $60$ nm were checked (see Fig.~S2 of the SM).

For all CG-model simulations, we set the depth of the potential well of the cross-interacting polymer beads to $\epsilon^{\rm B}=0.1$~kcal/mol and the reference temperature to $T = 300$~K. The superscript B refers to the polymer beads. When comparing the CG model against the all-atom MD modelling, the CG ions are set to mimic Na$^+$ and Cl$^-$ as hydrated ions.
Following Refs.~ \citenum{whitley2004,mag_cl,freeman2011}, the CG model cation (Na$^+$ equivalent) and anion (Cl$^-$ equivalent) are set to $\sigma^{\rm Na}=0.234$ nm and $\sigma^{\rm Cl}=0.378$ nm. The corresponding depths of the WCA potential wells are set to $\epsilon^{\rm Na}=1.3\epsilon^{\rm B}=0.13 \ {\rm kcal/mol}$ and $\epsilon^{\rm Cl}=1.24\epsilon^{\rm B}=0.124 \ {\rm kcal/mol}$, respectively. PE charge of $-74 e$ is neutralized by $74$ monovalent counter-cations. In analogy to atomistic-detail simulations, we examine the system response to added salt concentrations of $0.125$, $0.25$, $0.5$, and $1$ M. 

The CG-MD simulations are performed using the LAMMPS simulation package~\cite{plimpton1995,Thomson2022}. The long-range electrostatic interactions are calculated using the particle-particle particle-mesh method (PPPM)~\cite{plimpton1997}. 
A real space cutoff of $1.5$ nm is used, beyond which the contributions are obtained in reciprocal space. The PPPM relative force accuracy is set to $10^{-5}$. The $NVT$ ensemble is used, with the temperature of the system maintained at $300$ K using the Nose-Hoover thermostat~\cite{nose1984, hoover1985}, and with a coupling constant of $0.2$ ps. The rod beads are set immobile while the mobile ions are initially created at random non-overlapping positions using the Packmol package~\cite{packmol}.
The equations of motion for the mobile ions are integrated using a velocity Verlet algorithm with time step of $2$ fs. The production run duration is $8$ ns out of which the first $0.5$ ns are omitted in data analysis as initial equilibration time based on potential energy and CG ion distributions reaching equilibrium. For visualization, the VMD~\cite{vmd} and OVITO~\cite{ovito} packages were used.

\subsection{Poisson-Boltzmann theory}\label{sec:PB_theory}

Consider an infinite, impenetrable, rigid and charged cylinder with point-like neutralizing counterions with surface charge density of $\lambda/2\pi a$, where $a$ is the cylinder radius. The full nonlinear Poisson-Boltzmann (PB) equation describing the electric potential $\phi(r) = \phi_{\rm PB}(r)$ surrounding such a cylinder is given by \cite{grochowski2008}
\begin{equation}\label{nonlinear_pb2}
 \nabla^2 \phi_{\rm PB}(r)= -\frac{e}{\varepsilon} \mathlarger{\sum_{ i}} \zeta^{ i} \rho_0^{ i}  \exp\Big(-\beta e \zeta^{ i} \phi_{\rm PB}(r)\Big),
\end{equation}
where $\zeta^{i}$ is the ion valency and $\rho_0^{ i}$ the number density of the $i^{\rm th}$ ion species. We consider here the case of negatively charged PEs which have to be neutralized by adding positive counterions of number density $\rho^{\rm ci}$. After this, salt is added to the system. Thus, the number density of anions equals that of the added salt $\rho_0$. The number density of cations then equals $\rho_0 + \rho^{\rm ci}$.
For the present geometry, Eq.~(\ref{nonlinear_pb2}) has to be solved numerically (details below). The ion number density can be obtained from
\begin{equation}
\rho_{\rm PB}^{ i}(r)=\rho_0^{ i} \exp\Big(-\beta e \zeta^{ i} \phi_{\rm PB}(r)\Big).\label{density_theory2}
\end{equation}

\subsection{Linear Poisson-Boltzmann theory}\label{sec:lpb}

When $\beta e\zeta^{i}\phi_{\rm PB}(r)\ll1$, which corresponds to assuming small electrostatic potentials, one can linearize the full PB equation. This linear approximation is commonly referred to as Debye-H\"uckel approximation. Equation~(\ref{nonlinear_pb2}) then becomes
\begin{equation}\label{linear_pb2}
 \nabla^2 \phi_{\rm LPB}(r)=\kappa^2 \phi_{\rm LPB}(r),
\end{equation}
where
\begin{equation}\label{kappa_definition}
 \kappa^2=\frac{\sum_{i} \zeta^2e^2\rho^{i}_{0}}{\varepsilon k_{\rm B}T}.
\end{equation}

\noindent Here $1/\kappa$ is the Debye or screening length~\cite{grochowski2008}. A charged particle that is closer than $1/\kappa$ to the cylinder surface feels the surface charge density of the cylinder and thus interacts with it. On the other hand, a particle that find itself further than such distance is shielded from the intervening salt solution, which weakens the attraction or repulsion between the ion and the cylinder. The analytic solution of Eq.~(\ref{linear_pb2}) for $r>a_{\rm }$ is~\cite{brenner1974, trizac2006}
\begin{equation}
\phi_{\rm LPB}(r) =\frac{\lambda}{2\pi a\kappa \varepsilon }\frac{K_{0}(\kappa r)}{K_{1}(\kappa a_{\rm })},\label{eq:solP_1cyl}
\end{equation}
where $K_0$ and $K_1$ are the zeroth- and first-order modified Bessel functions of the second kind, respectively. The ion charge density can be calculated via Eq.~(\ref{density_theory2}) substituting $\phi_{\rm PB}$ with $\phi_{\rm LPB}$. 

\subsection{Soft-Potential-Enhanced  Poisson-Boltzmann theory}

Here we propose a modification to Eq.~(\ref{density_theory2}) as follows:
\begin{equation}
\rho^{i}_{\rm SPB/SLPB}(r)=\rho_0^{i} \exp\Big({-\beta\zeta^{i} e \phi_{\rm PB/LPB}(r)-\beta\tilde{V}^{i}_{\rm }(r)}\Big),\label{density_theory3}
\end{equation}
which replaces the impenetrable cylinder with a soft cylinder, and the potential $\phi(r)$ comes either from the full (PB) or linearized (LPB) theory, respectively. The modification facilitates a more realistic description of the ion densities of finite-sized ions close to the surface of the cylinder, and allows nonzero values for $\rho_{\rm SPB}(r)$ for $r < a$. 

To formulate soft-potential modification of Eq.~(\ref{density_theory3}), let us consider the repulsive interaction perceived by a single ion due all the $N$ coarse-grained polymer beads labeled as B in the system. The effective potential $V_{\rm }^{i}(\textbf{r}^{ i})$ felt by ion labeled as $i$ at $\textbf{r}^{i}$ can be obtained by summing over the individual potentials of individual beads via
 \begin{equation}
      V_{\rm }^{i}(\textbf{r}^{i})=\begin{cases}4\epsilon^{{\rm B} i}\mathlarger{\mathlarger{\sum_{k}}} \left[\left[\frac{\sigma^{{\rm B}i}}{r^{i}_k}\right]^{12} -\left[\frac{\sigma^{{\rm B}i}}{r^{i}_k} \right]^6 \right] 
     +\epsilon^{{{\rm B} i}},\> r^{i}_k<r_{\rm c}^{{\rm B}i};\\
     0, \>{\rm otherwise,}\end{cases}
 \label{sum1}
 \end{equation}
where $r^{i}_k=\mid\textbf{r}^{i}-\textbf{r}^{\rm B}_k\mid$, $\textbf{r}^{\rm B}_{k}$ is the position of the $k^{\rm th}$ bead, $\epsilon^{{\rm B}i}=\sqrt{\epsilon^{\rm B} \epsilon^{i}}$, and $r_{\rm c}^{{\rm B}i}$ is the corresponding cutoff distance. If the value of $b$ is small enough, one can fit Eq.~(\ref{sum1}) with a cylindrically symmetric WCA potential of the form
\begin{equation}
       \tilde{V}^{i}_{\rm}(r)=\begin{cases}4\tilde\epsilon^{{\rm B}i} \left[\left[\frac{\tilde \sigma^{{\rm B}i}}{r^{i}}\right]^{12} -\left[\frac{\tilde \sigma^{{{\rm B}i}}}{r^{i}}\right]^6 \right] 
     +\tilde\epsilon^{{{\rm B}i}}, \> r^{{\rm B}i}<\tilde r^{{\rm B}i}_{\rm c};\\
     0, \>{\rm otherwise,}\end{cases}
 \label{effect_WCA}   
\end{equation}
where $r^{i}$ is the radial distance from the center of the cylinder and the values of $\tilde\epsilon$, $\tilde \sigma$ are obtained numerically. The cutoff radii for the pair interactions are set to $\tilde r^{{\rm B}i}_{\rm c}=2^{1/6}\tilde \sigma^{{\rm B}i}$.

We note that in an earlier work, Heyda and Dzubiella~\cite{heyda} introduced an empirical correction to Eq.~(\ref{density_theory2}), where $\tilde{V}_{\rm }(r)$ is assumed to be of Gaussian form with three fitting parameters. As can be seen in Refs.~\onlinecite{heyda} and \onlinecite{batys2017}, this correction is successful in capturing chemically specific, finite-sized ion distributions in many cases. The correction of our work is motivated by the physical properties of the CG simulations model and requires only one parameter whose value turns out to be robust for a wide range of salt and ion sizes as demonstrated in this work.

When the soft-potential Poisson-Boltzmann model is compared against the CG model in terms of ability to capture ion size influence, only the counterions (monovalent cations) are present.
Equation (\ref{nonlinear_pb2}) is solved numerically using the finite-element calculation package of COMSOL Multiphysics\textsuperscript{\textregistered} software v5.2, COMSOL Inc., with boundary conditions $\phi_{\rm }(\infty)=0$ and $\phi'_{\rm}(a)= \lambda/2\pi a\varepsilon$, where $\phi_{\rm}'$ is the first spatial derivative. Where relevant, the modification of Eq.~(\ref{density_theory3}) is considered. We have carefully benchmarked the numerics against known results to guarantee convergence of the solutions.

\section{Results}\label{sec:results_discussion}

To benchmark the CG model and our soft-potential-enhanced Poisson-Boltzmann (SPB) theory against a chemically specific PE with atomistic details, we consider first the atomistic model of the PSS molecule in explicit water in varying monovalent salt NaCl concentrations. Additionally, the atomistic-detail simulations provide the coarse-grained model a diameter for the polymer beads $\sigma^{\rm B}$ in the potential of Eq.~(\ref{effect_WCA}). The soft-potential-modified LPB and PB radii $a$ (defined in Sec.~\ref{sec:PB_theory}) are optimized against the ion distributions from both the atomistic and CG molecular dynamics simulations.

\begin{figure*}[t!]
 \centering
 \includegraphics[width=1\linewidth]{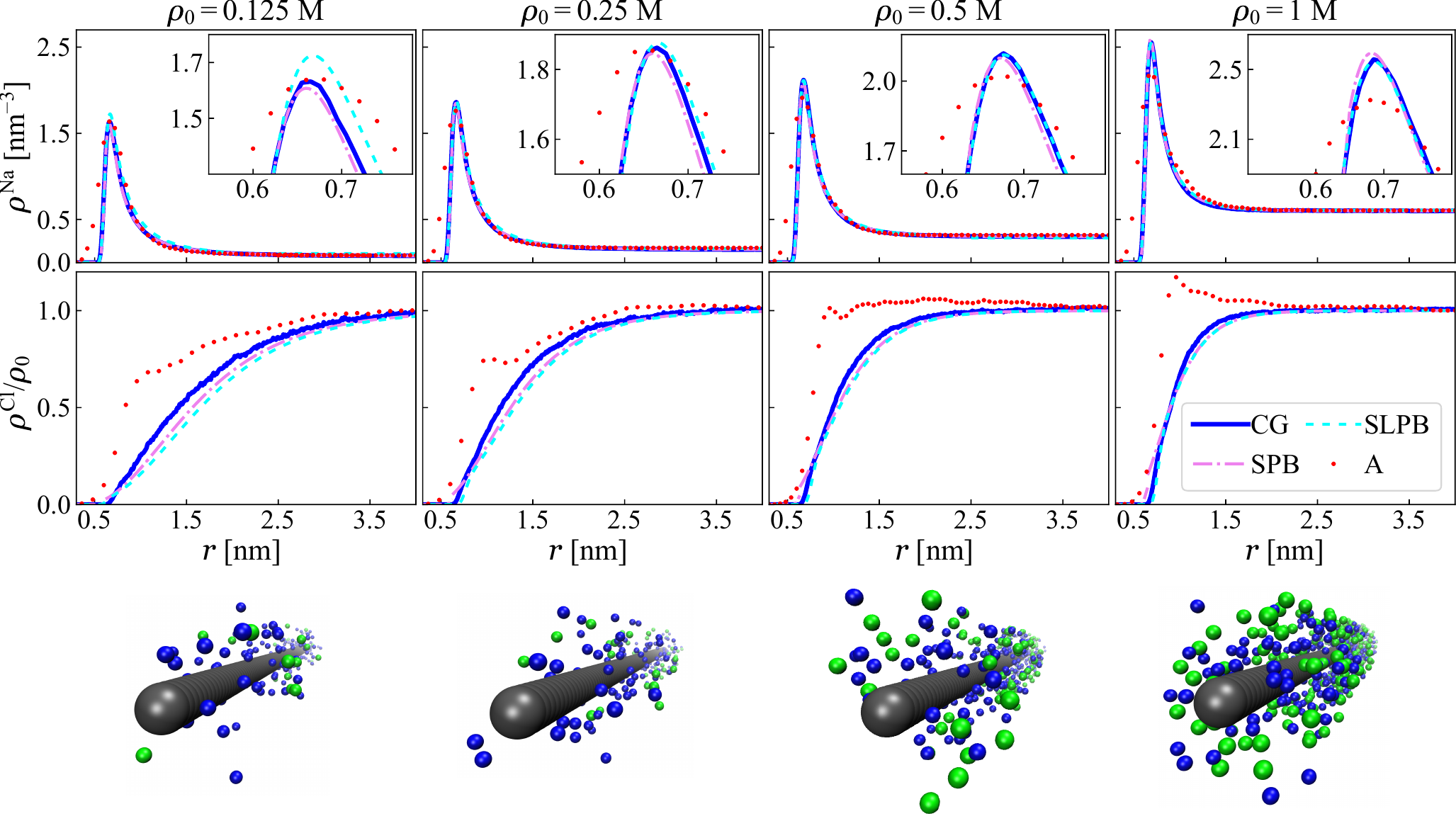}
 \caption{Na$^+$ (upper panels) and Cl$^-$ ion (lower panels) mean number density vs. radial distance $r$ from the PE axis for atomistic-detail PSS (A, dotted red line), soft-radius-optimized CG model (CG, solid blue line), full SPB theory (SPB, dash-dot purple line) and SLPB theory (SLPB, dashed cyan line) for different added NaCl concentrations $\rho_0$. The linear charge density of the polymer is $\lambda =-3.6$  e/nm. The snapshots are from the coarse-grained model system after a $8$ ns of MD simulation. CG polymer beads are in grey, Na${^+}$ in blue and Cl${^-}$ in green, respectively.}
  \label{salt_pss}
\end{figure*}

First, the CG-model polymer bead diameter $\sigma^{\rm B}$ is optimized such that the number density profiles of the Na$^+$ ions of the CG model, $\rho^{\rm Na}_{\rm CG}$, approximate the ones obtained from atomistic MD simulations $\rho^{\rm Na}_{\rm A}$. This is done such that the optimal polymer bead diameter $\sigma^{\rm  B}_{\rm *}$ minimizes the root-mean-square error (RMSE) between the atomistic modeling and the CG modelling Na$^+$ ion mean radial densities measured from the PE center of mass, see SM for technical details.
An optimal diameter $\sigma^{\rm B}_{*}=1.08$ nm is found at a salt concentration of $\rho_{0}=0.125$ M.  Most importantly, we varied the salt concentration up to $1$~M and found that $\sigma^{\rm B}_{*}$ remained constant to a good degree of approximation. Thus, no refitting of this soft radius is needed for the entire range of physical conditions studied in this paper, demonstrating the robustness of our approach. Ions with diameter $\sigma^{ i}$ of $0.1$, $0.2$, $0.5$, and $1.08$ nm are examined. For all cases, we use $\tilde{\sigma}^{{\rm B}i}=\sigma^{{\rm B}i}$. The parameter $\tilde\epsilon^{{\rm B}i}$ is found to be $0.07$, $0.08$, $0.11$ and $0.135$ kcal/mol, respectively. For the case of Na$^+$ and Cl$^-$ ions, we set $\tilde{\sigma}^{\rm B Na}=0.66$ nm and $\tilde{\sigma}^{\rm B Cl}=0.729$ nm, and $\tilde\epsilon^{\rm BNa}=0.107$ kcal/mol and $\tilde\epsilon^{\rm B Cl}=0.11$ kcal/mol, respectively.

To obtain accurate results from the SPB and SLPB theories, the corresponding cylinder radii $a_{\rm SPB}$ and $a_{\rm SLPB}$ have to be adjusted in Eq. (\ref{density_theory3}) for the electrostatic potential $\phi(r)$.
Unlike the robust soft-cylinder radius, these effective radii may change with system conditions. The optimization is done by comparison of the numerical Na$^+$ ion densities from both the CG and the atomistic models with the SPB and SLPB theories of Eq.~(\ref{density_theory3}) while keeping
the soft-cylinder diameter fixed at $\sigma^{\rm B}_*$.
Based on the ion densities, Eqs.~(\ref{nonlinear_pb2}) and (\ref{linear_pb2}) allow us to calculate $\phi_{\rm PB}(r)$ and $\phi_{\rm LPB}(r)$ for different values of $a$. We identify the optimal radii $a^*_{\rm SPB}$ and $a^*_{\rm SLPB}$ giving rise to the optimal potentials $\phi_{\rm SPB}^*(r)$ and $\phi_{\rm SLPB}^*(r)$ in Eq. (\ref{density_theory3}) based on minimizing the RMSE between $\rho_{\rm CG}^{\rm Na}$ and $\rho_{\rm SPB}^{\rm Na}$ (or $\rho_{\rm SLPB}^{\rm Na}$). Additional information about the RMSE calculation is provided in SM. The SLPB and SPB optimized radii at different salt concentrations are summarized in Table~\ref{opt_tab}. It is interesting to note that for a different salt, the radius from the SPB theory does not change.

Next, we turn to explore the range at which the CG model is able to capture the atomistic ion distribution and the applicability of the SPB model to reproduce both the atomistic and the CG ion distributions. The focus is on the PE in salt concentrations ranging from counterions only to $1$ M added monovalent salt. First, 
Fig.~\ref{salt_pss} shows Na$^+$ and Cl$^-$ ion number density profiles at different salt concentrations from atomistic and CG MD as well as the SLPB and SPB models. Additionally, CG modelling snapshots of the simulation region close to the polymer at equilibrium ion distribution (after $8$ ns) are presented. The data show that the CG model with optimized and fixed $\sigma^{\rm  B}_{*}$ approximates the atomistic simulations for Na$^+$ number density profiles around the PSS very well, except for the shortest distances from the polymer axis where the atomistic structure of the PSS allows higher ion densities than the CG model. It is also notable that the charge in the atomistic-detail PSS is located at the sulfonate groups at the ends of the side chain functional groups. The effect of this in the ion distribution can be seen in Fig.~S3 of the SM, which shows by 2D ion distribution the radial inhomogeneity of the atomistic ion distribution. Figure S3 of the SM presents also the CG 2D ion distributions. The radially symmetric CG model captures the mean density of the ion distribution. We note that the SPB theory faithfully follows the CG data for salt up to $1$ M: this ability of the model to follow accurately the spatial dependence of the radially symmetric ion densities from the CG model should be emphasized. 

\begin{figure*}[t!]
 \centering
     \includegraphics[width=0.95\linewidth]{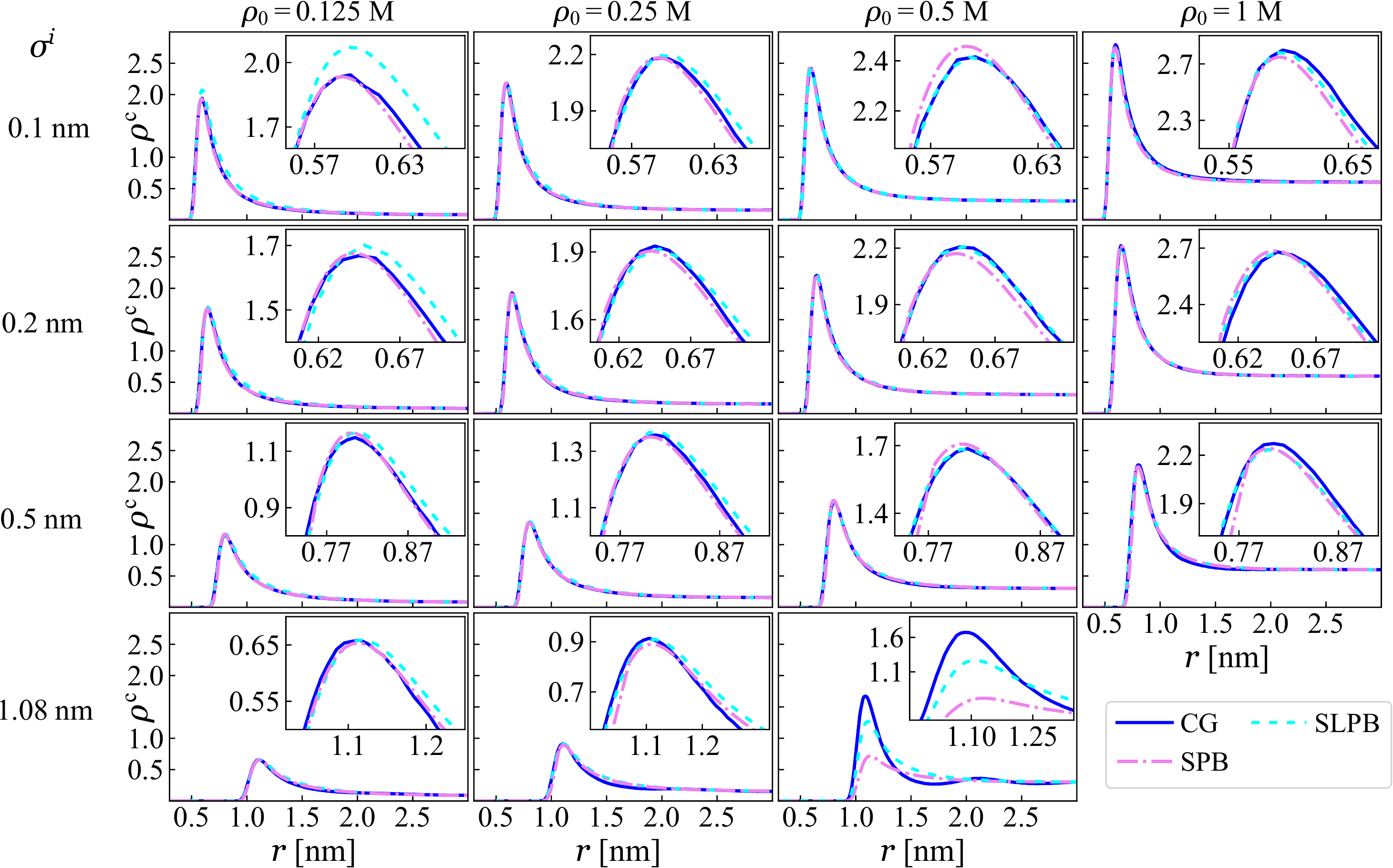}
\caption{Radial cation number densities $\rho^{\rm c}$ (in units of nm$^{-3}$) for varying bulk salt concentrations $\rho_0$ and different ion diameters $\sigma^{i}$. For each case, we compare the results of the CG-MD simulations with SLPB and SPB theory
 and the insets show details of the data close to the peaks. The linear charge density of the polymers is $\lambda =-3.6$  e/nm. The case $1$ M, $1.08$ nm is reported in Fig. S4 of the SM.}
  \label{ion_size_salt_concentration}
\end{figure*}

To check for the numerical stability of the ion distribution in the CG model, we tested different exponents in the interactions potential Eq.~(\ref{lj1}), including the $9-6$ potential. The ion distribution is relatively insensitive to the precise form of the potential, given that an attractive well is included.

\subsection{Ion-size effects}

The absolute values of $\phi_{\rm PB}$ and $\rho_{0}$ are inversely proportional. This is due to screening effect perceived by ions when they are further away from the polymer. Higher salt concentrations produce a stronger screening effect, which weakens the electrostatic interactions~\cite{grochowski2008}.
At the salt concentrations studied here, the screening effect is dominant and thus nonlinear effects covered by PB theory can be neglected. Consequently, the LPB (Debye-H\"uckel theory) remains an accurate approximation and successful in describing the potential and ion distribution. However, at low salt concentrations, nonlinear effects are important and the Debye-H\"uckel theory fails~\cite{klein2002, netz2003}.
As expected, e.g. in Refs.~\onlinecite{fogolari1999, deserno_and_barbosa2000, vieillefosse1981} it has been shown that LPB is efficient only at sufficiently high salt where the potentials remain small as compared to $k_{\rm B}T$.
A consequence is the limitation of the applicability of the LPB theory at low salt and for highly charged polymers~\cite{fogolari1999, 2015accuracy}.

\begin{table}[h!]
\caption{Summary of CG model parameters optimized based on the atomistic simulations. The data present for each salt ion concentration $\rho_0$, the radial distance of the  Na$^+$ ion number density peak from PSS axis $a_{\rm A}$, the interaction parameter between Na$^+$ ions and CG polymer beads $\sigma^{\rm BNa}$ (calculated by mixing rule), as well as, the optimized cylinder radii in SLPB model $a^*_{\rm SLPB}$ and SPB model $a^*_{\rm SPB}$.}
\label{opt_tab}
\begin{ruledtabular}
\begin{center}
\begin{tabular}{c  c c c c} 

$\rho_{0}$[M] & $a_{\rm A}$[nm] & $\sigma^{\rm BNa} $[nm] & $a^*_{\rm SLPB}$[nm] & $a^*_{\rm SPB}$[nm] \\ [0.5ex] 
 \hline
 $0.125$ & $0.68$ & $0.66$ & $0.24$ & $0.62$  \\ 

 $0.25$ & $0.68$ & $0.66$ & $0.46$ & $0.62$ \\

 $0.5$ & $0.68$ & $0.66$ & $0.57$ & $0.62$ \\

 $1$ & $0.68$ & $0.66$ & $0.62$ & $0.62$ \\
\end{tabular}
\end{center}

\end{ruledtabular}
\end{table}

\begin{figure*}[t!]
 \centering
 \includegraphics[width=1\linewidth]{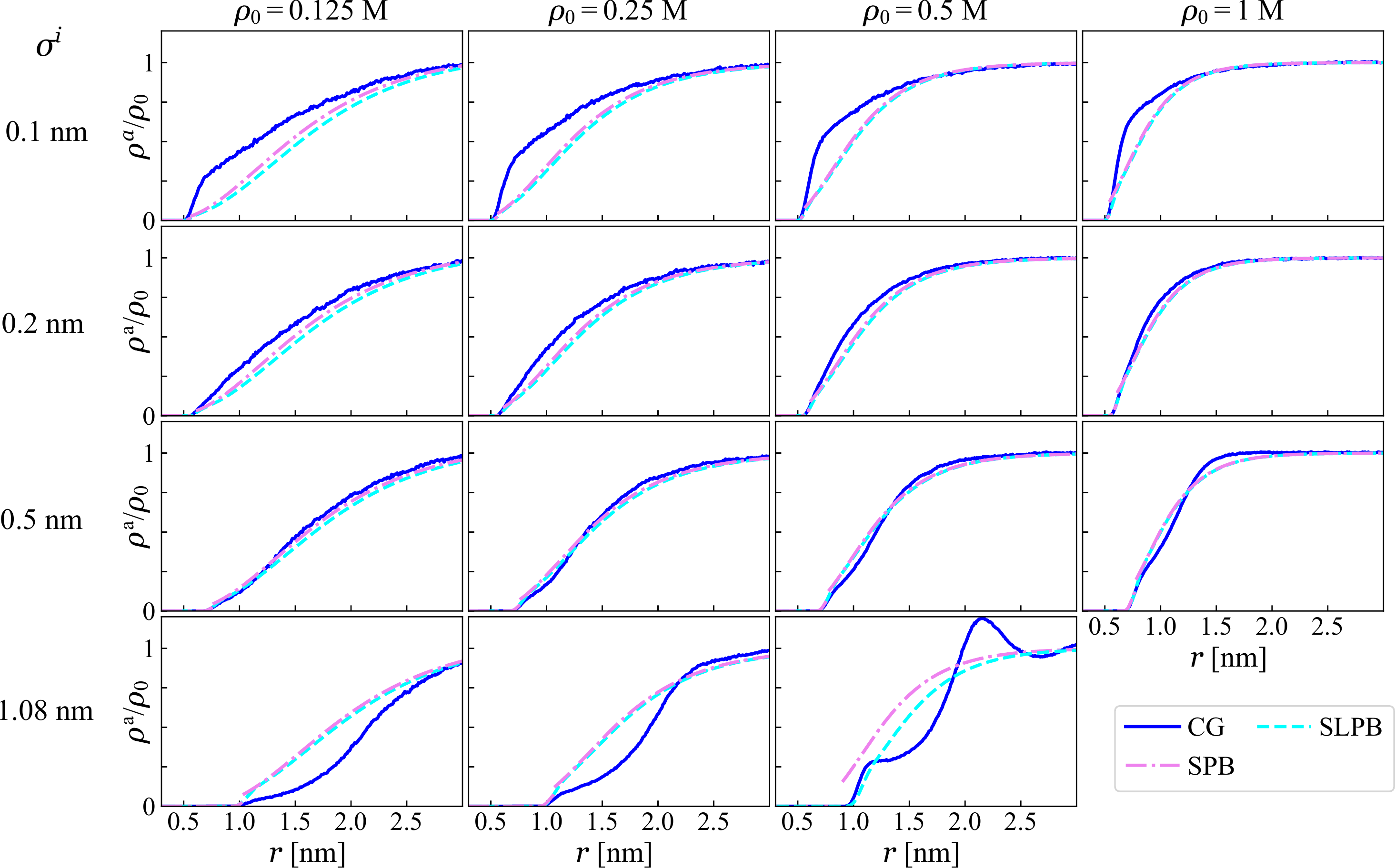}
 \caption{Normalized radial anion number densities $\rho^{\rm a}/\rho_0$ 
 for varying bulk salt concentrations $\rho_0$ 
 and ion diameters $\sigma^{i}$. 
 For each case we compare the results of the CG-MD simulations with SLPB and SPB theories. The linear charge density of the polymers is $\lambda =-3.6$ e/nm.
 The case $1$ M, $1.08$ nm is reported in Fig. S4 of the SM.}
  \label{Cl)16_densities} 
\end{figure*}

To examine the effect of ion size, we carried out CG-MD simulations at added salt concentrations of $0.125$, $0.25$, $0.5$, and $1$~M of monovalent $1:1$ salt where both the cation and anion have identical diameters of $0.1$, $0.2$, $0.5$ or $1.08$ nm, where the largest value corresponds to that of the optimized CG polymer bead diameter $\sigma^{\rm B}_*$. The ion number density distributions are calculated as before from Eq. (\ref{density_theory3}). 

Table~\ref{opt_tab_all_system} presents, for different added salt ion concentrations and diameters, the SLPB and SPB model optimal radii $a^*_{\rm SLPB}$ and $a^*_{\rm SPB}$ minimizing the RMSE between the cation number density profile obtained from the CG model, as well as the Manning radius $r_{\rm M}$~\cite{bret1984} and fraction of condensed counterions $x^{\rm c}_{\rm M}(r)$, based on the Manning-Oosawa condensation model~\cite{manning1969, oosawa1968}. The Manning radius $r_{\rm M}$ defines the shell corresponding to fraction of condensed counterions $1-1/\zeta_{\rm M}$~\cite{deserno2000paper}.
We determine $r_{\rm M}$ by the inflection point criteria~\cite{deserno2000paper} and $x^{\rm c}_{\rm M}(r)$ was calculated following Ref.~\onlinecite{batys2017}. 

In Figs.~\ref{ion_size_salt_concentration} and \ref{Cl)16_densities} we report the resulting cation and anion number densities, respectively. The SPB theory and the CG-MD model predictions agree very well except for the highest excess salt concentration $\rho_{0}=1$ M and the largest ion diameter $\sigma^{ i}=1.08$ nm. Data for this case are available in the SM. In this case, significant deviations, in practice positional correlations, are induced by steric packing considerations rising from the size and elevated concentration of the ions. Such steric and positional correlations due to the finite size and strong electrostatic correlations result in layered ion organization in the vicinity of highly charged PEs~\cite{colla2017}. 
Moreover, correlations between ions have also been suggested to induce additional attractive forces not captured by the PB theory~{\cite{ murthy1985, mills1985, pack1990, hecht1995, tomac1998, hatlo2009}}.
As a result of both the size~\cite{lamm1994} and this effect, the main density profile peak is underestimated by the SPB theory (see e.g. Fig.~\ref{ion_size_salt_concentration} for ion diameter of $1.08$ nm and salt concentration of $0.5$ M).
Beyond this any mean-field approach can be expected to fail when charge correlations become significant. This occurs for example for multivalent ion species~\cite{netz2000,chu2007}, localized charges~\cite{chu2007}, or large ion sizes~\cite{borukhov1997, li2009}.
It is worth mentioning that previous successful attempts to consider the finite size of ions in PB type models exist~\cite{borukhov1997, borukhov2000, colla2017,batys2017}.
We expect that the SPB theory could be extended by incorporating such modifications to better describe large ions.

Additionally, for small ion diameters at low salt concentrations, some significant deviations between the LPB theory and CG-MD simulations ion distributions rise. This is because the value of $\beta e\zeta^{ i}\phi_{\rm}(r)$ cannot be considered small. In fact, the LPB model predictions approach those of the PB theory only in the dense limit, in which the potentials remain small and thus the Debye-H\"uckel approximation is valid.

\begin{figure}[b!]
    \centering
    \includegraphics[width=0.95\linewidth]{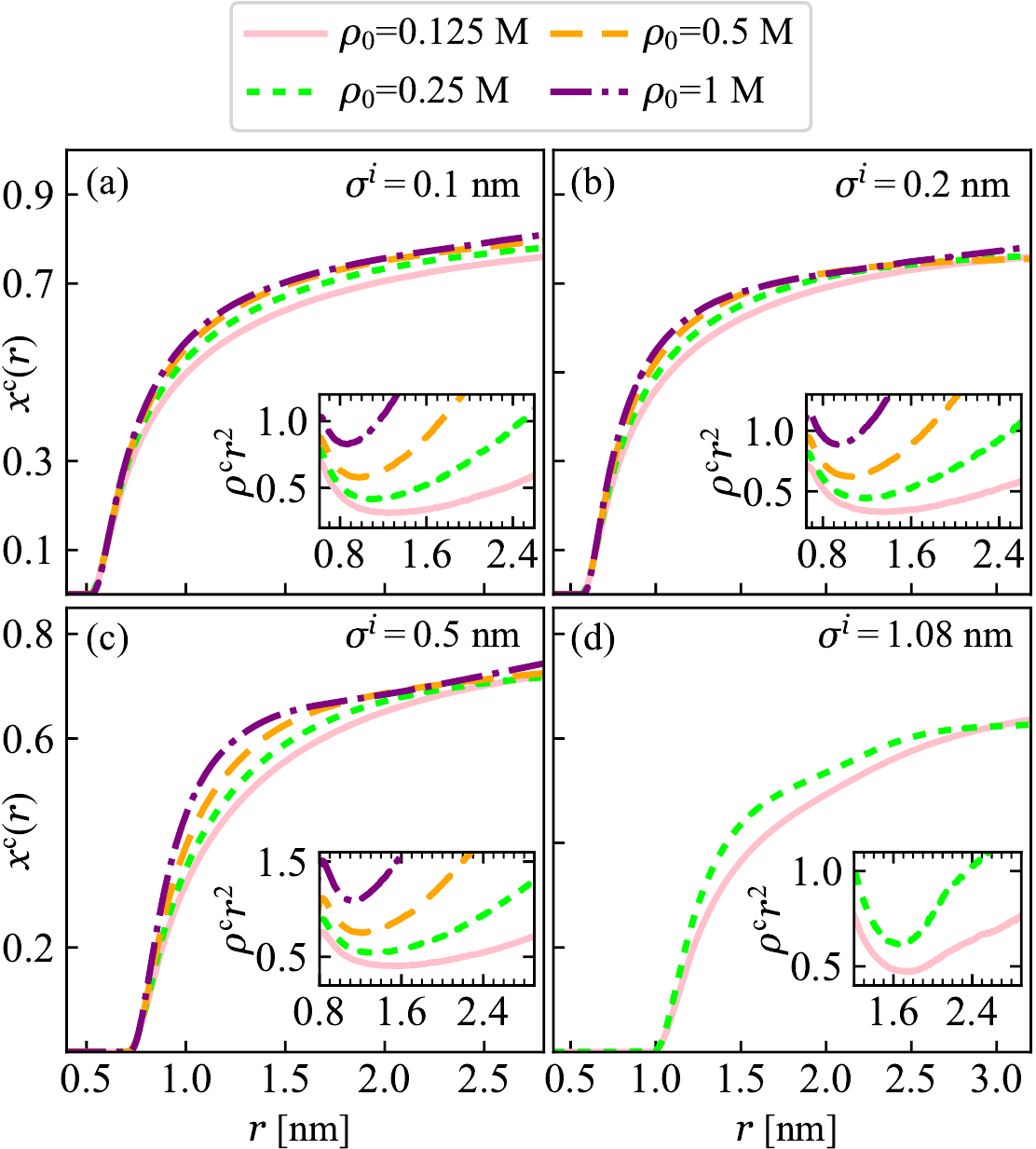}
    \caption{Fraction of counterions $x^{\rm c}(r)$ at a distance $r$ from CG polymer axis for different added salt concentrations. The ion diameters $\sigma^{i}$ are respectively (a) $0.1$ nm, (b) $0.2$ nm, (c) $0.5$ nm and (d) $1.08$ nm. The insets show $\rho^{\rm c}(r)r^2$ as a function of $r$. The Manning radii are the location of minima of the inset curves and are reported in Table~\ref{opt_tab_all_system}.}
    \label{fig_inflection_fraction}
\end{figure}

Let us next consider ion condensation. Table~\ref{opt_tab_all_system} shows the dependence of the Manning condensation fraction $x^{\rm c}_{\rm M}$ for the different ion diameters and concentrations. Figure~\ref{fig_inflection_fraction} presents the corresponding data such that the fraction of cations is plotted versus the distance $r$ from the PE. The inflection points defining the Manning radii are the minima of the curves in the insets of Fig.~\ref{fig_inflection_fraction}.
For systems without additional salt, $x^{\rm c}_{\rm min}=1-1/\zeta_{\rm M}$ represents the lower limit of condensed ion fraction~\cite{heyda}. For our system, this value is $x^{\rm c}_{\rm min}=0.60$, calculation based on Bjerrum length $\ell_{\rm B}=0.7$ nm in water at $300$ K and polymer axial charge density $\lambda=-3.6$ e/nm (Manning parameter $\zeta_{\rm M}=2.52$).

\begin{table}[hbt!]
\caption{Parameters for the interaction between ions and CG polymer beads $\sigma^{{\rm B}i}$, the optimized radii of hard cylinders in the SLPB ($a^*_{\rm SLPB}$) and the SPB ($a^*_{\rm SPB}$) models, Manning radii ($r_{\rm M}$), and the fraction of condensed ions ($x^{\rm c}_{\rm M}$), respectively.}
\label{opt_tab_all_system}
\begin{ruledtabular}
\begin{center}
\begin{tabular}{c c c c c  c c } 

$\rho_{0}$[M] & $\sigma^{i}$[nm] & $\sigma^{{\rm B}i}$[nm] & $a^*_{\rm SLPB}$[nm] & $a^*_{\rm SPB}$[nm] & $r_{\rm M}$[nm] & $x_{\rm M}^{\rm c}$ \\ [0.5ex] 
 \hline
 $0.125$ & $0.1$ & $0.59$ & $0.08$ & $0.56$ & $1.235$ & $0.58$  \\ 
 
 $0.25$ & $0.1$ & $0.59$ & $0.36$ & $0.56$ & $1.105$ & $0.57$ \\
 
 $0.5$ & $0.1$ & $0.59$ & $0.46$ & $0.56$ & $0.985$ & $0.54$ \\
 
 $1$ & $0.1$ & $0.59$ & $0.5$ & $0.54$ & $0.855$ & $0.46$ \\
 
 $0.125$ & $0.2$ & $0.64$ & $0.18$ & $0.61$ & $1.365$ & $0.60$  \\ 
 
 $0.25$ & $0.2$ & $0.64$ & $0.43$ & $0.61$ & $1.185$ & $0.57$ \\
 
 $0.5$ & $0.2$ & $0.64$ & $0.53$ & $0.61$  & $1.065$ & $0.55$\\

 $1$ & $0.2$ & $0.64$ & $0.58$ & $0.61$ & $0.925$ & $0.48$ \\
 
 $0.125$ & $0.5$ & $0.79$ & $0.45$ & $0.76$ & $1.485$ & $0.55$ \\ 
 
 $0.25$ & $0.5$ & $0.79$ & $0.63$ & $0.76$ & $1.305$ & $0.52$\\
 
 $0.5$ & $0.5$ & $0.79$ & $0.72$ & $0.77$ & $1.225$ & $0.52$ \\
 
 $1$ & $0.5$ & $0.79$ & $0.76$ & $0.78$ & $1.155$ & $0.52$\\
 
 $0.125$ & $1.08$ & $1.08$ & $0.82$ & $1.03$ & $1.755$ & $0.43$\\ 
 
 $0.25$ & $1.08$ & $1.08$ & $1.0$ & $1.07$ & $1.645$ & $0.46$ \\
 
 $0.5$ & $1.08$ & $1.08$ & $1.08$ & $1.08$ & $-$ & $-$ \\
 
 $1$ & $1.08$ & $1.08$ & $1.08$ & $-$ & $-$ & $-$ \\
\end{tabular}
\end{center}
\end{ruledtabular}
\end{table}

The data of Table~\ref{opt_tab_all_system} show that the fraction of condensed ions indeed approaches the predicted minimum for small ion diameters. Additionally, for fixed ion diameter, a smaller condensation is predicted as the excess salt concentration is increased (except for $\sigma^{ i}=1.08$ nm, where the large radius induces inaccuracies even for those systems that exhibit an inflection point). The explanation of this trend is twofold: ($i$) Manning condensation theory involves point-like counterions and thus does not account for finite ion sizes~\cite{manning1969, oosawa1968}, implying that it is expected to work better for smallest ions; ($ii$)
$x^{\rm c}_{\rm min}$ is calculated for a system with counterions only. Since here excess ions are added in the system, the ionic screening occurs at smaller distances and shifts $r_{\rm M}$ to smaller values. Consequently, also the value of $x^{\rm c}_{\rm M}$ decreases. Notably, particle-based simulations have reported condensation below the Manning prediction several times before~\cite{antila2014,savelyev2006,antila2015b}. 
For chemically specific interactions models, the prediction is naturally subject to ion models employed~\cite{yoo2012}.

\begin{figure}[b]
 \centering
     \includegraphics[width=0.98\linewidth]{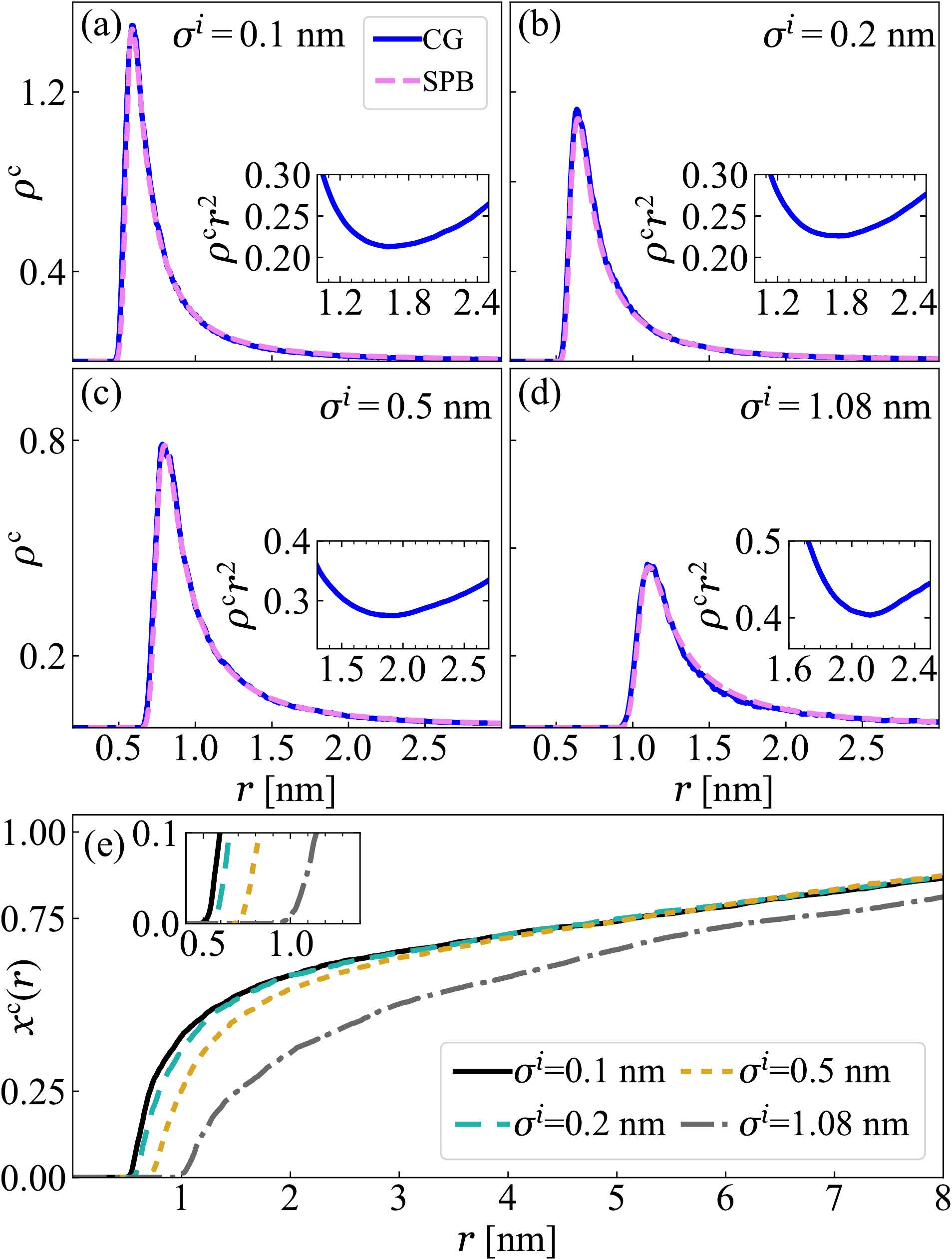}
 \caption{Comparison of the radial cation densities $\rho^{\rm c}$ [nm$^{-3}$] for ion diameters $\sigma^{i}$ (a) $0.1$ nm, (b) $0.2$ nm, (c) $0.5$ nm, and (d) $1.08$ nm from the CG-MD simulations and the SPB theory in the case of counterions only, i.e. no excess salt in the systems. Note that the linear charge density of the polymer is $\lambda =-3.6$  e/nm. Insets show $\rho^{\rm c}(r)r^2$ as a function of distance $r$ from the polymer axis. Panel (e) presents CG simulation results for the ion-size dependence of the  fraction of counterions $x^{\rm c}$ at distance $r$. The inset shows a zoom of the condensation-onset part of the data.}
  \label{salt-free_density}
\end{figure}

\subsection{Salt-free systems}
As the nonlinear effects in the PB models become pronounced in the low-salt limit, we next examine the influence of ion diameter in the absence of added salt, i.e. in systems with only counter-cations present. 
The optimal radii $a^*_{\rm SPB}$ that minimize the RMSE between the cation number density distribution of the CG model and the one obtained from the SPB theory are listed in Table~\ref{opt_tab_saltfree} for ions with diameters of $0.1 $, $0.2 $, $0.5$, and $1.08$ nm. Figure~\ref{salt-free_density} shows the corresponding cation number density profiles. The SLPB theory predictions are omitted here, since it completely fails to capture the electrostatic potential due to the low bulk density of cations. Thus, its ability to model the ion number density distribution is poor. 
The number density distributions reveal that the enhanced SPB theory is able to correctly capture the position and the magnitude of the condensation peak for small and intermediate ion diameters. However, the agreement worsens when the ion size becomes comparable to the polymer diameter.

Figure~\ref{salt-free_density}(e) shows the ion size dependence of the fraction of counterions around the PE as a function of the radial distance from its center $x^{\rm c}(r)$. The curves are similar, but, as expected, are differentiated by a shift towards larger values of $r$ as the ion diameter is increased. Additionally, for large ions, some underlying structure due to packing effects can be seen.     
Table~\ref{opt_tab_saltfree} shows the dependence of the corresponding $x^{\rm c}_{\rm M}$ values with respect to the ion diameter. The fraction of condensed ions decreases slightly when the ion diameter is increased.

\begin{table}[h!]
\begin{ruledtabular}
\caption{Summary of parameters for different ion diameters in a system where no added salt is present (countercations only).  Parameters for the interaction between the cations and the CG polymer beads $\sigma^{{\rm B}i}$, optimized radii for the SPB theory $a^*_{\rm SPB}$, the Manning radii $r_{\rm M}$, and the fraction of condensed ions $x_{\rm M}^{\rm c}$ are presented. }
\label{opt_tab_saltfree}
\begin{center}
\begin{tabular}{ c  c  c c  c } 
$\sigma^{ i}$ [nm] & $\sigma^{{\rm B}i} $ [nm] & $a^*_{\rm SPB}$ [nm] & $r_{\rm M}$ [nm] & $x_{\rm M}^{\rm c}$ \\ [0.5ex] 
 \hline
 $0.1$ & $0.59$ & $0.57$ &$1.6$ & $0.67$ \\ 

 $0.2$ & $0.64$ & $0.59$ &$1.73$ & $0.67$ \\
 
 $0.5$ & $0.79$ & $0.76$ & $1.92$ & $0.65$ \\

 $1.08$ & $1.08$ & $1.07$ & $2.13$ & $0.48$ \\

\end{tabular}
\end{center}
\end{ruledtabular}
\end{table}

\section{Summary and Conclusions}\label{sec:conclusion}

We have presented a conceptually simple soft-potential-enhanced Poisson-Boltzmann theory based on a CG-particle-based model for PEs in monovalent ion solutions. The findings demonstrate that our simple modification to the PB equations enables the PB approach to accurately predict spatial ion distributions for a single rodlike PE under a large variety of ionic conditions, including salt concentrations up to $1$~M and a variety of ion diameters ranging from those corresponding to small, hard monovalent ions such as Na or Cl to significantly larger ions.  

The CG model is optimized via a single, physically motivated and robust fitting parameter which corresponds to the effective soft diameter of the CG polymer. Here, the optimization was based on atomistic-level simulations of PSS polyelectrolyte for a fixed salt concentration, but similar approach can be used for diameters otherwise obtained. The diameter was found to be virtually independent of the added salt concentration up to at least $1$ M monovalent salt and for a relatively highly charged PE such as PSS. Even for an asymmetric PE such as PSS, a good agreement between the cation number density distributions obtained from the CG and the atomistic simulations was found.

Finally, we have addressed the performance of the approach when there is no added salt. The SPB theory was able to accurately capture the position and the magnitude of the condensation peak for small and intermediate ion sizes. Also, the fraction of condensed cations was addressed and compared with the predictions from Manning theory. 
Our work provides an easy way to implement significant enhancement to the standard PB theory for description of PEs in aqueous salt solutions.
In future work we will apply the theory to study PE-PE interactions in aqueous solutions.

There are a number of interesting additional research questions that could be studied using the SPB approach developed here. The fundamental limitation of the theory comes from the weak-coupling and mean-field nature of the PB equation. Multivalent ions induce charge correlations that cannot be captured by theories based on the PB equation and thus we have not considered them here. Ion-shape effects that induce additional steric interactions could be included in the SPB approach and work in this direction is in progress. We have also assumed throughout the work that the PB equation has strict radial symmetry, i.e. the PE is a rigid rod. Capturing flexibility and local curvature of real PEs would be complicated. The resulting complex boundary conditions pose a significant challenge for solving the full (non-linear) PB equation {\cite{forster1995}}. Furthermore, possible sequence heterogeneity significantly changes the local ion condensation and induces spatial correlations that are also beyond the standard PB theory where an average surface charge is assumed~{\cite{eggen2009}}.

\section*{SUPPLEMENTARY MATERIAL}
This Supplementary Material consist of data demonstrating the degree of smoothness of equipotential lines around the polyelectrolyte rod in the coarse-grained (CG) molecular dynamics (MD) simulations, finite-size effects, the two-dimensional ion number density distribution maps of the atomistic and CG molecular dynamics simulations, ion number density profiles corresponding to the large, $1.08$~nm ion diameter at $1$~M salt concentration in the CG-MD simulations, details on the root-mean-square error (RMSE) analysis, a summary of simulation system details, and an input script file for the CG-MD simulations as well as an extract from the initial configuration data file.

\section*{Acknowledgements}
This work was supported by the Academy of Finland through its Centres of Excellence Programme (2022-2029, LIBER) under project no. 346111 (M.S.) and Academy of Finland projects Nos. 309324 (M.S.) and  307806 (PolyDyna) (T.A-N. and X.Y.). The work was also supported by Technology Industries of Finland Centennial Foundation TT2020 grant (T.A-N. and X.Y.).
 We are grateful for the support by FinnCERES Materials Bioeconomy Ecosystem. Computational resources by CSC IT Centre for Finland and RAMI -- RawMatters Finland Infrastructure are also gratefully acknowledged. 

\section*{Data Availability}
Data to reproduce the figures in the manuscript is provided at
\href{https://doi.org/10.24342/a86e0ced-f0d6-4c0a-a10d-e87070b8569d}{https://doi.org/10.24342/a86e0ced-f0d6-4c0a-a10d-e87070b8569d}.
An example of the job script input file and the initial structure
file for CG-MD simulations are available within the supplementary material of DOI: \href{https://doi.org/10.1063/5.0092273}{10.1063/5.0092273}. If using the inputs or the open data, we request
acknowledging the authors by a citation to the original source  DOI: \href{https://doi.org/10.1063/5.0092273}{10.1063/5.0092273}.

\bibliographystyle{ieeetr}

\bibliography{refs}

 \pagebreak

\onecolumngrid
\begin{center}
{\Large \bf Supplementary Material}
\end{center}
\vspace{8mm}
\twocolumngrid

\setcounter{equation}{0}
\setcounter{table}{0}
\setcounter{figure}{0}

\renewcommand{\theequation}{S\arabic{equation}}
\let\oldthetable\thetable
\renewcommand{\thetable}{S\oldthetable}
\renewcommand{\thefigure}{S\arabic{figure}}

\section*{Quantification of equipotential lines smoothness in the coarse-grained molecular dynamics simulations}\label{app:c}

The total potential energy that Na$^+$ ions experience in the CG-MD simulations is obtained by summing the contributions of the $N$ beads in the rigid chain. In the simulations, the beads forming the polyelectrolyte (PE) rod are fixed in position with inter-bead spacing such that the counterions feel an effectively smooth surface potential. Data for this are shown in Fig.~S2. The figure shows equipotential lines for different values of the distinct and total interaction contributions over a $z$-axis segment of the PE rod. The data for the different energy contributions are presented for the bead-spacing parameter $b=0.27$ nm. 

\section*{Finite-size effects}
\begin{figure}[htb!]
 \centering
 \includegraphics[width=1\linewidth]{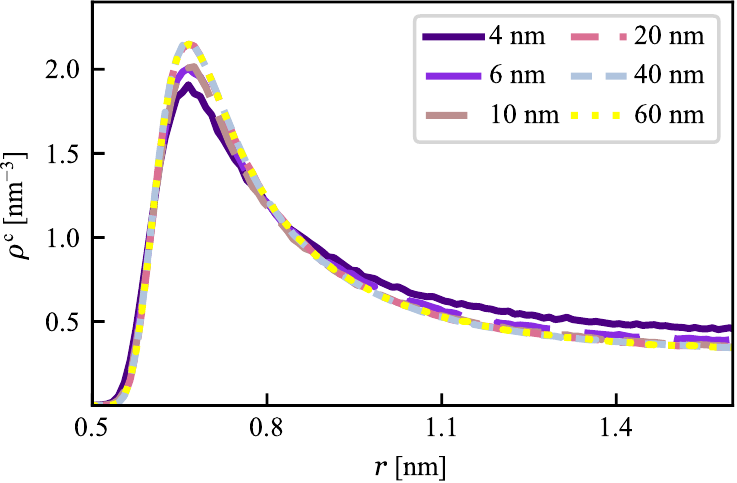}
 \caption{Na$^+$ ion mean number density vs. radial distance $r$ from the PE axis for the soft-radius-optimized CG model for cubic simulation boxes with sides between $4$ nm and $60$ nm. The linear charge density of the polymer is $\lambda =-3.6$  e/nm and the salt concentration is $0.5$ M.}
 \label{finite_size}
\end{figure}

Finite size effects were examined by comparing the Na$^+$ ion condensation from CG-MD simulations. Cubic boxes of sides $4$, $6$, $10$, $20$, $40$, and $60$ nm were checked. In Fig.~{\ref{finite_size}} we show the Na$^+$ number density profiles, converging for cubic simulation box sizes of ($20$ nm)$^3$ or larger.

\begin{figure*}[b]
 \centering
 \includegraphics[width=1\linewidth]{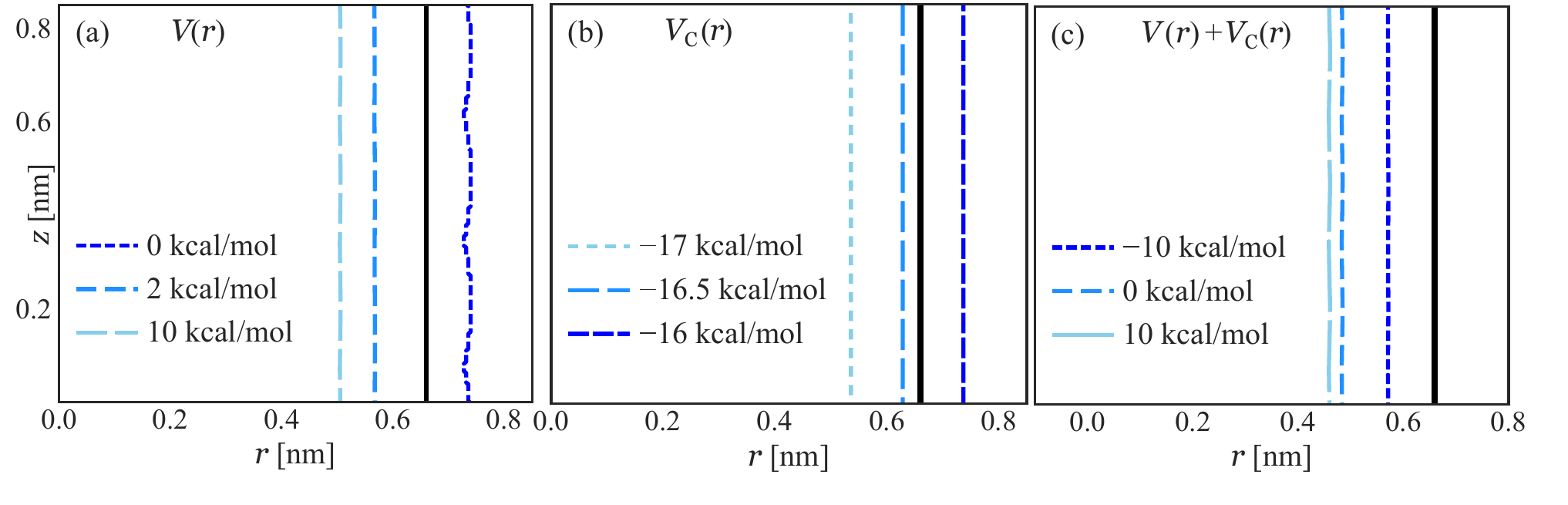}
 \caption{Equipotential lines of from the CG model as a function of radial distance $r$ from the center of the PE. Data are presented for (a) the Weeks-Chandler-Andersen (WCA) potential $V(r)$~(Eq.~(1) of the main text), (b) Coulombic potential $V_{\rm{C}}(r)$~(Eq.~(2), main text), and (c) the total potential energy $V(r)+V_{\rm{C}}(r)$ acting on Na$^+$ ions.
 The black solid lines correspond to $\sigma^{\rm BNa}$.
 The PE orientation follows Fig. 1 of the main text (polymer backbone along the $z$ axis).}
 \label{fig:equipotential}
\end{figure*}

\section*{Root mean square error (RMSE) analysis}\label{app:B}
The root-mean-square error (RMSE) between two quantities $\rho_1(x)$ and $\rho_2(x)$ is defined as
\begin{equation}
{\rm RMSE} =\frac{1}{\sqrt{N}}\sqrt{\sum_{k=1}^N \mid \rho_{1}(x_k)-\rho_{2}(x_k)\mid^2},
\label{rms_pss} 
\end{equation}
where $N$ is the number of sampling points.

\section*{Density maps}\label{app:D}
Figure S3 presents a comparison of the ion distributions in the atomistic-detail simulations and the CG-MD simulations as a two-dimensional density map. The data are averaged along the $z$ axis to obtain the $xy$-plane-average ion number densities at different salt concentrations.  Due to the asymmetric nature of PSS, the atomistic simulations show spatial variation (radial dependence) in the ion distribution, as expected. However, the CG model leads to a symmetric result because of the perfectly cylindrical geometry of the CG PE. Nevertheless, the degree of ion condensation is in good agreement between the two descriptions, especially when considering the radial average.
\begin{figure*}[bht!]\label{fig:append_2d}
 \centering
 \includegraphics[width=1\linewidth]{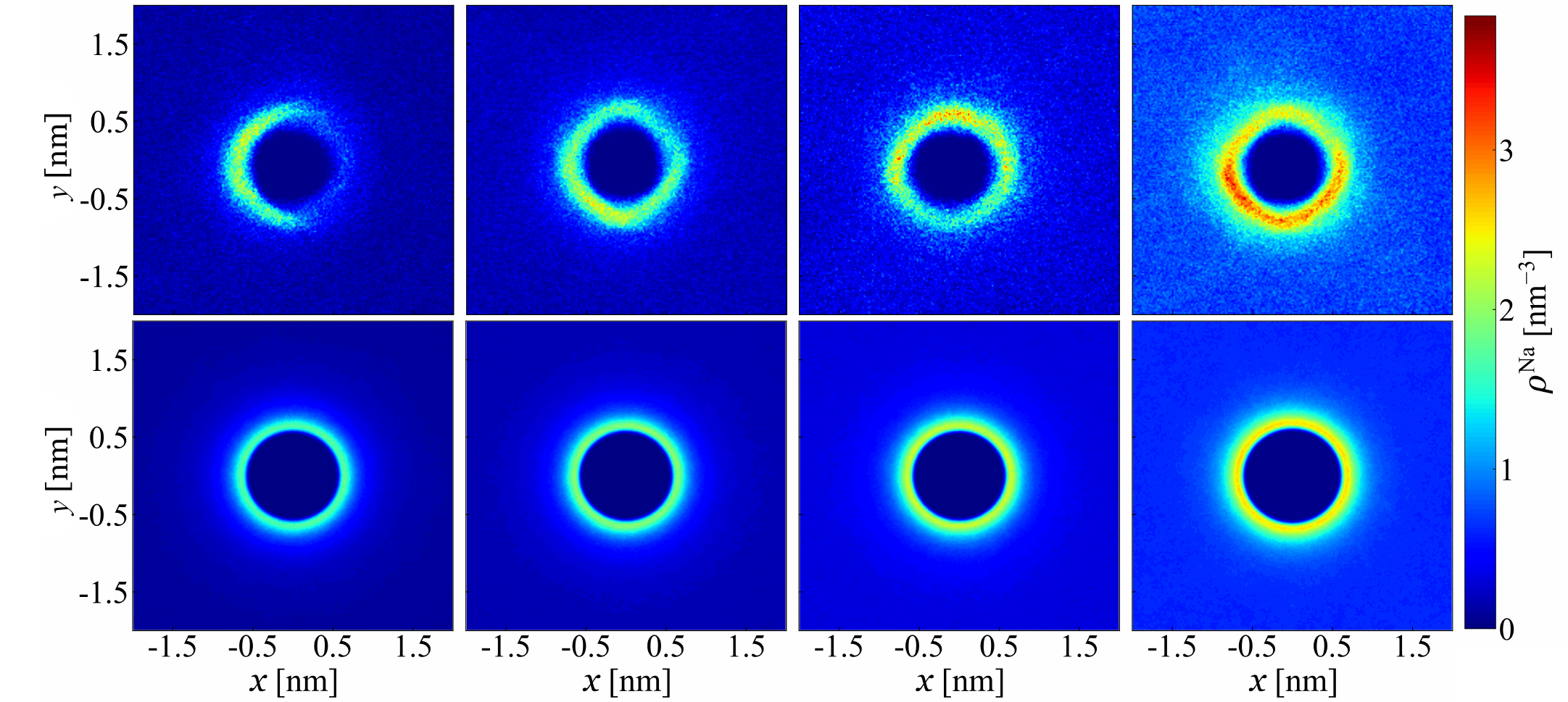}
  \caption{ Comparison of two-dimensional Na$^+$ ion average density distributions from atomistic detail PSS (top row) and optimized CG model (bottom row) simulations at excess NaCl concentrations $0.125$ M, $0.250$ M, $0.5$ M, and $1$ M (from left to right). The data are averaged along the $z$ axis. The atomistic-detail PSS and the CG polymer are at the center of the simulations box, aligned along the $z$ axis.}
 \label{2d_pss}
\end{figure*}

\begin{figure*}[htbp!]
 \centering
 \includegraphics[width=0.6\linewidth]{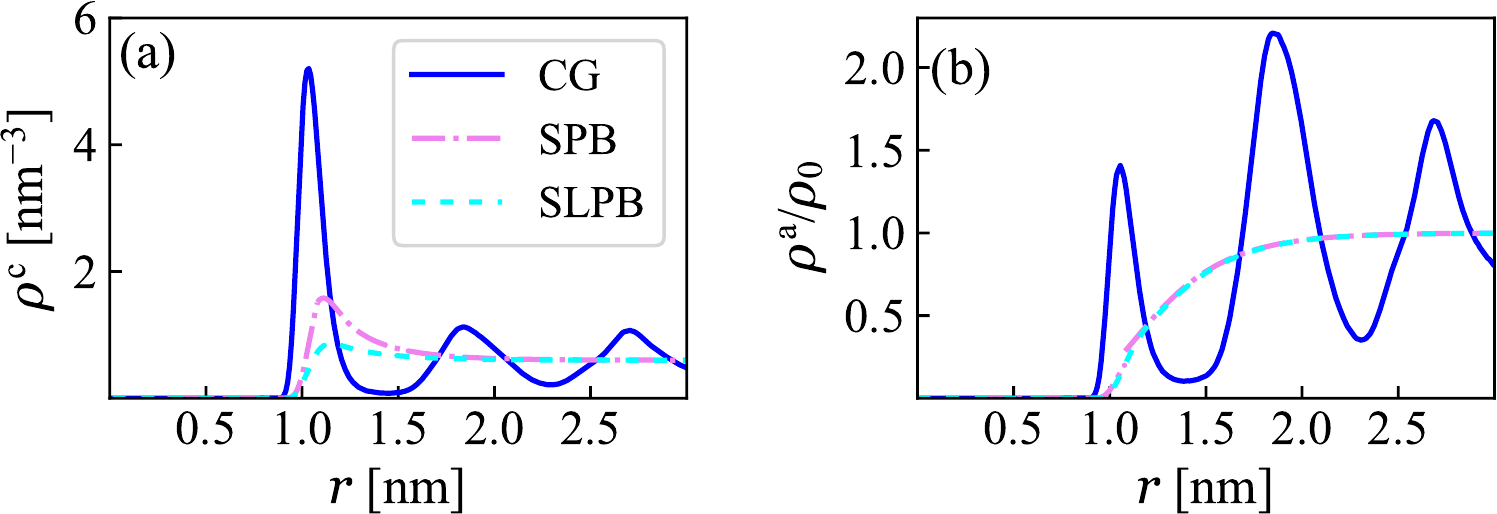}
 \caption{Radial (a) cation and (b) anion number densities $\rho(r)$ for salt concentration of $1$ M and ion diameters of $1.08$ nm (both anion and cation). Data from the CG-MD simulations and SPB and SLPB theories are presented.}
 \label{bigions_highsalt}
\end{figure*}

\section*{Data for ion distribution corresponding to large ion diameter at high salt concentration }\label{app:A}
In Fig.~S4, we compare the number density profiles of cations and anions from the full soft-potential-enhanced Poisson-Boltzmann (SPB) theory, its linearized version (SLPB), and the CG model. The ions and the beads of the polymers have the same diameter equal to $1.08$ nm, and the concentration of the monovalent salt  is $\rho_0=1$ M. The packing effects of the bulky large-radius ions lead to a layered ion structure around the PE. The layering and positional correlations driven by the steric packing constraints can be seen from the CG-MD model results. The PB theory fails to describe this case, since the approximation of pointlike ions does not hold in this scenario.


\section*{Summary of simulation details}\label{app:E}

In Tables SI and SII,  we present the dimensions of the simulation boxes and the number of water molecules and ions in each system for the atomistic-detail MD and the CG-MD simulations, respectively.

\begin{table}[thpb!]
\caption{Atomistic-detail MD simulation system details. The table presents for different excess salt concentrations $\rho_0$ the simulation box Cartesian dimensions $L_{x}$, $L_{y}$, and $L_{z}$, the number of water molecules $N^{\rm H_2O}$, the number of Na$^+$ ions $N^{\rm Na}$, and the number of Cl$^-$ ions $N^{\rm Cl}$.  }
\label{table_atomistic_SI}
\begin{ruledtabular}
\begin{center}
\begin{tabular}{c  c c c c c c} 

$\rho_0$ [M] & $L_{x}$ [nm]& $L_{y}$ [nm] & $L_{z}$ [nm] & $N^{\rm H_2O}$ & $N^{\rm Na}$  & $N^{\rm Cl}$ \\ [0.5ex] 
 \hline
 $0.125$ & $7.91$ & $7.91$ & $5.5$ & $11292$ & $46$ & $26$ \\ 
 
$0.25$ & $7.89$ & $7.89$ & $5.5$ & $11240$ & $72$ & $52$ \\

$0.5$ & $7.87$ & $7.87$ & $5.5$ & $11136$ & $124$ & $104$ \\

 $1$ & $7.84$ & $7.84$ & $5.5$ & $10928$ & $228$ & $208$ \\
 
\end{tabular}
\end{center}
\end{ruledtabular}
\end{table}

\begin{table}[hbtp!]
\caption{Coarse-grained MD simulation system details. The table presents for different salt concentrations $\rho_0$ the  simulation box Cartesian dimensions  $L_{ x}$, $L_{y}$, and $L_{z}$, the number of Na$^+$ and  Cl$^-$ ions $N^{\rm Na}$ and $N^{\rm Cl}$, respectively.}
\label{table_CG_SI}
\begin{ruledtabular}
\begin{center}
\begin{tabular}{c  c c c c c c} 
$\rho_0$ [M] & $L_{x}$ [nm]& $L_{y}$ [nm] & $L_{z}$ [nm] & $N^{\rm Na}$  & $N^{\rm Cl}$ \\ [0.5ex] 
 \hline
 $0.125$ & $20$ & $20$ & $20$ & $674$ & $600$ \\ 
 
$0.25$ & $20$ & $20$ & $20$ & $1274$ & $1200$ \\
 
$0.5$ & $20$ & $20$ & $20$ & $2474$ & $2400$ \\
 
 $1$ & $20$ & $20$ & $20$ & $4874$ & $4800$ \\
\end{tabular}
\end{center}
\end{ruledtabular}
\end{table}

\end{document}